\documentclass{jpsj-suppl}
\usepackage{txfonts}

\title{Pion Induced Reactions for the Study of Charmed Baryons}

\author{Sang-Ho \textsc{Kim}$^{1,2}$, Atsushi Hosaka$^{1}$, 
Hyun-Chul Kim$^{3}$, and Hiroyuki \textsc{Noumi}$^{1}$}

\inst{$^{1}$ Research Center for Nuclear Physics (RCNP), Osaka University,
Ibaraki, Osaka, 567-0047, Japan \\
$^{2}$ Asia Pacific Center for Theoretical Physics (APCTP), Pohang,
Gyeongbuk, 790-784, Republic of Korea \\
$^{3}$ Department of Physics, Inha University, Incheon 402-751, 
Republic of Korea}

\email{sangho.kim@apctp.org}

\recdate{October 23, 2015}

\abst{In this talk, we report the study of $\pi^- p \to K^{*0}\Lambda$ and 
$\pi^- p \to D^{*-}\Lambda_c^+$ reactions by using an effective Lagrangian 
method and a hybrid Regge model.
The total and differential cross sections for the $K^{*0}\Lambda$ 
production are first calculated and then those for the $D^{*-}\Lambda_c^+$ 
one are estimated.
The two models yield different results, each exhibiting specific features.
Our prediction for the charm production provides valuable information 
about the upcoming J-PARC experiment.}

\kword{charmed baryons, effective Lagrangians, Regge model}

\begin{document}
\maketitle

\section{Introduction}
The production of open charmed mesons and baryons is now a major  
issue in hadron physics, as the electromagnetic- and hadron-beam 
energies inrease enough to produce them at various experimental
facilities. For example, there is a plan at J-PARC to conduct experiments 
to investigate the charmed baryons via the pion-induced reactions at a
high-momentum beam line of up to 20 GeV/$c$~\cite{e50}. 
The only previous work regarding this process $\pi^- p \to D^{*-}B_c$, 
where  $B_c$ denotes a charmed baryon in ground or excited states 
($B_c=\Lambda_c^+, \Sigma_c^+,...$), was carried out almost thirty
years ago~\cite{Christenson:1985ms} at BNL. However, none of signals
for the charmed baryons was found but only an  
upper limit (95$\%$ confidence level) was estimated, namely, 7 nb at
13 GeV pion-beam energy. Thus it is of great importance to study
the production mechanism of this reaction systematically. 

On the theoretical side, the differential cross sections $d\sigma/dt$ for 
the strangeness and charm production, i.e. $\pi^- p \to K^{*0}\Lambda$ and 
$\pi^- p \to D^{*-}\Lambda_c^+$, have been computed with a simple Regge 
model~\cite{Kim:2014qha}, where vector-meson reggeon exchange was
considered to provide a rough estimate of the relative strength. 
However, we want to elaborate the study of Ref.~\cite{Kim:2014qha},
employing both an effective Lagrangian method and a hybrid Regge model.
We take into account the contribution of $K\,(D)$ and
$\Lambda\,(\Lambda_c)$ reggeons as well as that of $K^*\,(D^*)$
reggeon to the strangeness (charm) production. The Regge parameters
are fixed by using the quark-gluon string model (QGSM) done in
Ref.~\cite{Brisudova:2000ut}.   

\section{Effective Lagrangians}
\begin{figure}[tbh]
\centering
\includegraphics[width=9.0cm]{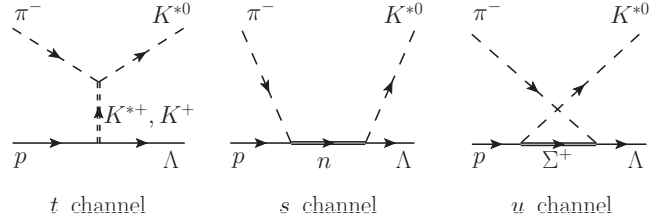}
\caption{Tree diagrams for the $\pi^- p \to K^{*0}\Lambda$ reaction.}
\label{FIG1}
\end{figure}

We start with the strangeness production process
$\pi^- p \to K^{*0}\Lambda$, for which the relevant tree-level Feynman
diagrams are depicted in Fig.~\ref{FIG1}.
We define the effective Lagrangians for each vertex by
\begin{eqnarray}
\mathcal L_{\pi K K^*} &=&
-ig_{\pi K K^*} ( \bar K \partial^\mu \tau \cdot \pi K^*_\mu - 
               \bar K_\mu^* \partial^\mu \tau \cdot \pi K ),           \cr
\mathcal L_{\pi K^* K^*} &=&
g_{\pi K^* K^*} \varepsilon^{\mu\nu\alpha\beta}
\partial_\mu \bar K_\nu^* \tau \cdot \pi \partial_\alpha K_\beta^*,
\end{eqnarray}
for the meson-meson-meson interactions, and
\begin{eqnarray}
\mathcal L_{K N \Lambda} &=&
\frac{g_{K N \Lambda}}{M_N+M_\Lambda} \bar N \gamma_\mu  
\gamma_5 \Lambda \partial^\mu K + \mathrm{H.c.},                      \cr
\mathcal L_{\pi N N} &=&
\frac{g_{\pi N N}}{2M_N} \bar N \gamma_\mu
\gamma_5 \partial^\mu \tau \cdot \pi N,                               \cr
\mathcal L_{\pi \Sigma \Lambda} &=&
\frac{g_{\pi \Sigma \Lambda}}{M_\Lambda+M_\Sigma} \bar \Lambda \gamma_\mu
\gamma_5 \partial^\mu \pi \cdot \Sigma + \mathrm{H.c.},               \cr   
\mathcal L_{K^* N Y} &=&                                                  
-g_{K^* N Y} \bar N \left[ \gamma_\mu Y - 
\frac{\kappa_{K^* N Y}}{M_N+M_Y} \sigma_{\mu\nu} Y 
\partial^\nu \right] K^{*\mu} + \mathrm{H.c.}.                         
\label{eq:Lag1}
\end{eqnarray}
for the meson-baryon-baryon interactions, $Y$ being the $\Lambda$ or
$\Sigma$ fields generically. The coupling constants are determined by
using the experimental data for 
hadron decays, Nijmegen-potential, or SU(3) flavor symmetry.
For more details, we refer to Ref.\cite{Kim:2015ita}.
From the Lagrangians given above, we can construct the invariant
amplitudes for each channel. Since the relevant hadrons have internal
structures, the following form factor is introduced to each amplitude:  
\begin{eqnarray}
F_{ex}(p^2) = \frac{\Lambda^4}{\Lambda^4+(p^2-M_{ex}^2)^2},
\label{eq:FF1}
\end{eqnarray}
where $p$ denotes the transfer momentum of the exchanged particle.
To get the amplitudes for the charm process, 
$\pi^- p \to D^{*-}\Lambda_c^+$, we just replace the strange mesons and
hyperons with charmed ones 
$K^+ \to \bar D^0,\, K^{*+} \to \bar D^{*0},\, \Lambda \to \Lambda_c^+,\,
\Sigma^+ \to \Sigma_c^{++}$.
Regarding the cutoff masses, we choose the the same values for the
$t$-channel and the $s$- and $u$- channels, respectively:  
$\Lambda_{K(D),\,K^*(D^*)}=0.55$ GeV,
$\Lambda_{N,\,\Sigma(\Sigma_c)}=0.60$ GeV. 

\begin{figure}[thp]
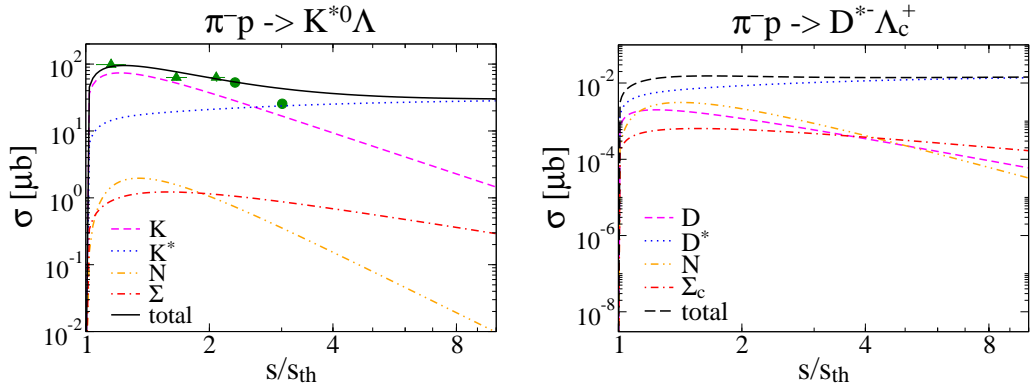

\centering
\includegraphics[width=6.5cm]{FIG2.eps} \hspace{1.0em}
\includegraphics[width=6.5cm]{FIG3.eps}
\caption{Total cross sections for the $\pi^- p \to K^{*0}\Lambda$ (left
panel) and $\pi^- p \to D^{*-}\Lambda_c^+$ (right panel) reactions based
on the effective Lagrangian method.
The data are from Ref.~\cite{Dahl:1967pg} (triangle) and 
from Ref.~\cite{Crennell:1972km} (circle).}
\label{FIG2}
\end{figure}
In Fig~\ref{FIG2}, the numerical results for the total cross sections are 
drawn as functions of $s/s_{\mathrm{th}}$, where $s_{\mathrm{th}}$ is the 
threshold of $s$.
In both reactions, we can find that the $t$-channel process gives the 
most dominant contribution to the total cross section.
Especially, vector-meson exchanges play important roles in the high 
energy region.

\begin{figure}[thp]
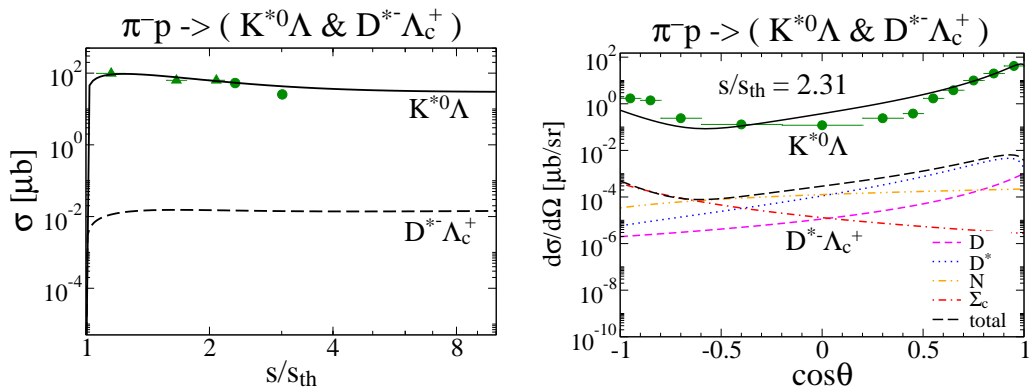

\centering
\includegraphics[width=6.5cm]{FIG4.eps} \hspace{1.0em}
\includegraphics[width=6.5cm]{FIG5.eps}
\caption{Total (left panel) and differential (right panel) cross sections 
for the $\pi^- p \to K^{*0}\Lambda$ reaction in comparison with the
$\pi^- p \to D^{*-}\Lambda_c^+$ one based on the effective Lagrangian 
method.
The data are from Ref.~\cite{Dahl:1967pg} (triangle) and 
from Ref.~\cite{Crennell:1972km} (circle).}
\label{FIG3}
\end{figure}
Figure~\ref{FIG3} describes the difference between the strangeness and
charm production.
It turns out that the cross sections for the charm production are 
approximately $10^4$ times smaller than that for the strangeness one.  
Because of the large energy scale for the charm production compared to
the strangeness one, the Feynman propagators and form factors are the
main reason for this suppression.

\section{Regge model}
The Regge amplitudes are derived by replacing the Feynman propagator 
$P^{\mathrm{F}}$ with the Regge propagator $P^{\mathrm{R}}$
\begin{eqnarray} 
P_{K^*}^{\mathrm{F}} = \frac{1}{t-M_{K^*}^2} \Rightarrow
P_{K^*}^{\mathrm{R}}(s,t) = 
\left( \frac{s}{s_{K^*}} \right)^{\alpha_{K^*}(t)-1}
\Gamma[1-\alpha_{K^*}(t)] \alpha'_{K^*},                           
\label{EQ:RProp}
\end{eqnarray}
for $K^*$ reggeon exchange, for example.
To extract the Regge trajectory
$\alpha_{K^*}(t) = \alpha_{K^*}(0) + \alpha'_{K^*} t$ and scale parameter
$s_{K^*}$, we rely on the quark-gluon string model 
(QGSM)~\cite{Brisudova:2000ut}.
The following form factor is considered to each amplitude:
\begin{eqnarray}
F_{ex}(p^2) = \frac{a}{(1 - p^2/\Lambda^2)^2}.
\end{eqnarray}
The typical values are used for the cutoff masses 
$\Lambda = 1\,\mathrm{GeV}$ in common.
Other free parameters are chosen as 
$a_{K(D)}=0.6,\,a_{K^*(D^*)}=0.8$ and $a_{\Sigma(\Sigma_c)}=1.5$~\cite{Kim:2015ita}.

\begin{figure}[thp]
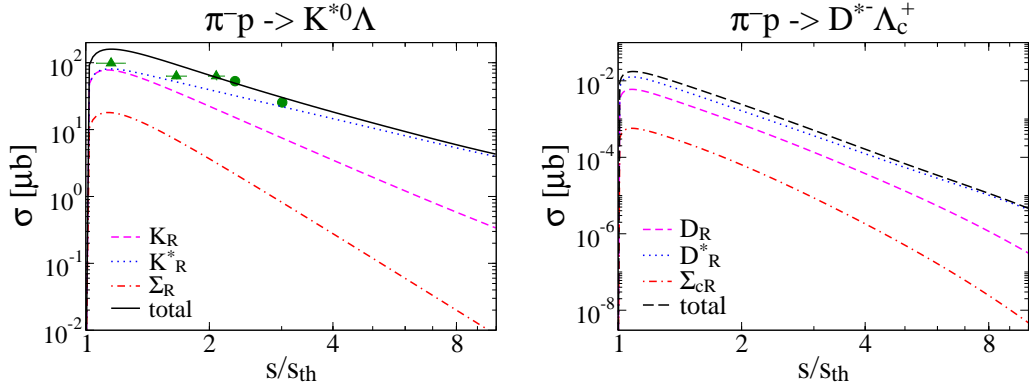

\centering
\includegraphics[width=6.5cm]{FIG6.eps} \hspace{1.0em}
\includegraphics[width=6.5cm]{FIG7.eps}
\caption{Total cross sections for the $\pi^- p \to K^{*0}\Lambda$ (left
panel) and $\pi^- p \to D^{*-}\Lambda_c^+$ (right panel) reactions based
on the hybrid Regge model.
The data are from Ref.~\cite{Dahl:1967pg} (triangle) and 
from Ref.~\cite{Crennell:1972km} (circle).}
\label{FIG4}
\end{figure}
In Fig~\ref{FIG4}, each contribution to the total cross section is
displayed.
In both reactions, vector-meson reggeon exchange ($K^*$ and $D^*$)  
governs its dependence on $s$ over the whole energy region.
The difference between pseudoscalar reggeon exchange and vector-meson
reggeon one gets larger as $s/s_{\mathrm{th}}$ increases. 
The reason is obvious from the values of 
$\alpha(0)$~\cite{Brisudova:2000ut}.
The contribution of baryon reggeon exchange ($\Sigma$ and $\Sigma_c$)
is almost negligible.

\begin{figure}[thp]
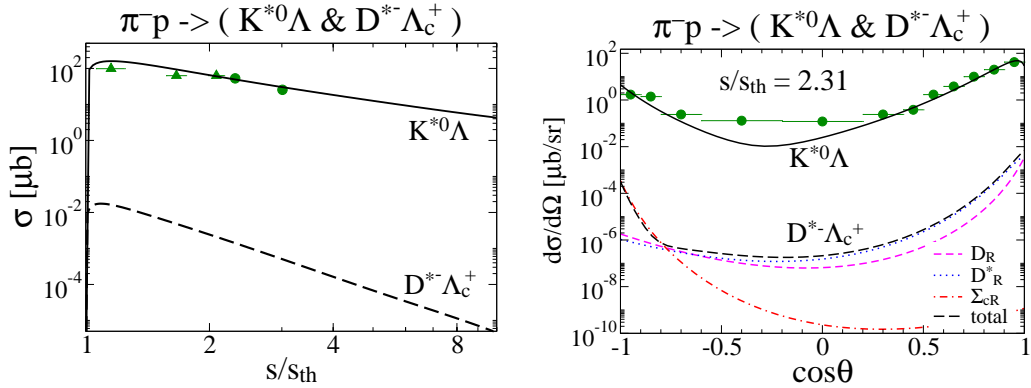

\centering
\includegraphics[width=6.5cm]{FIG8.eps} \hspace{1.0em}
\includegraphics[width=6.5cm]{FIG9.eps}
\caption{Total (left panel) and differential (right panel) cross sections 
for the $\pi^- p \to K^{*0}\Lambda$ reaction in comparison with the
$\pi^- p \to D^{*-}\Lambda_c^+$ one based on the hybrid Regge model.
The data are from Ref.~\cite{Dahl:1967pg} (triangle) and 
from Ref.~\cite{Crennell:1972km} (circle).}
\label{FIG5}
\end{figure}
In Fig.~\ref{FIG5}, the difference between the strangeness and charm
production is drawn.
It is found that the cross sections for the charm production are about 
$10^4 - 10^6$ times smaller than that of the strangeness production 
depending on the energy range.

\section{Summary}
In the present talk, we investigated the pion-induced $K^{*0}\Lambda$ and
$D^{*-}\Lambda_c^+$ production off the nucleon.
In general, vector-meson (vector-meson reggeon) exchange turns out to be 
a dominant contribution in the effective Lagrangian method (Regge 
model).
Out prediction of the charm production at
$s/s_{\mathrm{th}} \sim 2.1$, which is the
expected maximum energy from the J-PARC facility~\cite{e50}, is suppressed
by about factor $10^4$ compared to the strange production.  
This indicates that the production cross section for the 
$D^{*-}\Lambda_c^+$ reaction is around 2 $nb$ at that energy.

\section*{Acknowledgments}
This work is supported in part by the Grant-in-Aid for Science Research 
(C) 26400273.  
S.H.K. is supported by a Scholarship from the Ministry of Education, 
Culture, Science and Technology of Japan.
The work of H.-Ch.K. was supported by Basic Science Research Program 
through the National Research Foundation of Korea funded by the Ministry 
of Education, Science and Technology (Grant Number: 2013S1A2A2035612).
It is also supported in part by the NRF grant funded by MEST
(Center for Korean J-PARC Users, Grant No. NRF-2013K1A3A7A06056592).

\end{document}